\documentclass[twocolumn,prb,aps,showpacs,amsmath,amssymb]{revtex4}
\usepackage{amsfonts}
\usepackage{bbm}
\usepackage{graphicx}
\usepackage{dcolumn}
\usepackage{bm}
\newcommand\mycite[1]{\raisebox{-0.5em}{\Large\cite{#1}}}
\usepackage[bookmarks=true,colorlinks=false,hyperindex=true,pdfstartview=FitH]{hyperref}
\begin{document}
\title{Josephson junction on one edge of a two dimensional topological insulator affected by magnetic impurity}

\author{Shu-feng Zhang}
\author{Wei Zhu}
\author{Qing-feng Sun}
\email{sunqf@iphy.ac.cn}
\affiliation{Institute of Physics, Chinese Academy of Sciences, Beijing 100190, China}

\begin{abstract}
Current-phase relation in a Josephson junction formed by putting two
s-wave superconductors on the same edge of a two dimensional
topological insulator is investigated. We consider the case that the
junction length is finite and magnetic impurity exists. The similarity and
difference with conventional Josephson junction is discussed. The
current is calculated in the semiconductor picture. Both the $2\pi$-
and $4\pi$-period current-phase relations ($I_{2\pi}(\phi),
I_{4\pi}(\phi)$) are studied. There is a sharp jump at $\phi=\pi$
and $\phi=2\pi$ for $I_{2\pi}$ and $I_{4\pi}$ respectively in the
clean junction. For $I_{2\pi}$, the sharp jump is robust against
impurity strength and distribution. However for $I_{4\pi}$, the
impurity makes the jump at $\phi=2\pi$ smooth. The critical (maximum) current of $I_{2\pi}$ is given and we find it will be
increased by asymmetrical distribution of impurity.
\end{abstract}


\maketitle

\section{Introduction}

Recently the topological insulator (TI) has excited great interest in
the condensed-matter
community.\cite{review(Ti)(S.C.Zhang),review(Ti)(C.L.Kane)} The
unique feature of TI is the existence of edge states (or surface
states) which is protected by time reversal symmetry. The edge state
of a two dimensional (2D) TI can be considered approximately as a 1D
mental. But since spin and momentum direction of carriers is locked
together owing to strong spin-orbit coupling, it's only half of the
ordinary electron gas. This helical property is robust against
nonmagnetic impurity due to its topological origin. If the edge
state is in contact with a superconductor, a topological
superconducting edge state will form in the interface because of
proximity effect.
\cite{{PRL100-096407(2008)(L.Fu)},PhysRevB.79.161408(2009)(L.Fu)}
And it can be viewed as a 1D topological superconductor (TS).
Therefore it's able to construct a Josephson junction on one edge of
the 2D TI.

Experimentally, the edge state in HgTe/CdTe quantum
wells,\cite{science 318-766(2007)} in InAs/GaSb quantum wells,
\cite{PRL107-136603(2011)(RRDu)} and surface state in $Bi_2Se_3$
systems \cite{NatPhys5-398(2009)(Y.Xia)} have been observed. The
superconducting proximity effect and Andreev reflection in InAs/GaSb
quantum wells and $Bi_2Se_3$ systems coupling to superconducting
electrode have been demonstrated.
\cite{PRL109-186603(2012)(RRDu),Science336-52(2012)(Q.K.Xue)}

The conventional superconductor-normal metal-superconductor (SNS)
junction has been investigated in detail in the last three decades.
\cite{JETP30-944(1970)(Kulik),PhysRevB.46.12573(1992)(P.F.Bagwell),PhysRevLett75-1831(1995)(D.Averin),PhysRevB56-11232(1997)(M.Hurd),RevModPhys76-411(2004)(A.A.Golubov)}
Since the superconductor-TI-superconductor (STiS) junction is only
half of the SNS junction, the corresponding Andreev bound
state\cite{JETP30-944(1970)(Kulik)} and current-phase relation
are similar for the clean junction if we suppose quasiparticles
distribute thermodynamically ($2\pi$-period current
case).\cite{txt1} However for STiS junction, a $4\pi$-period
current-phase ($I_{4\pi}(\phi)$) relation (fractional Josephson
effect) may arise if the thermodynamical distribution is partially
destroyed while superconducting phase difference is changed
adiabatically. \cite{PhysRevB.79.161408(2009)(L.Fu),Kitaev1}
The effect of nonmagnetic impurity and magnetic impurity is
identical for the SNS junction due to spin degeneracy. However for the STiS
junction only magnetic impurity can lead to a backscattering owing
to time reversal symmetry. In dirty junctions magnetic impurity contributes another significant difference, the
extra $\pi$ phase shift for hole reflection.\cite{PRL107-177002(2011)}
As a result even the $2\pi$-period
current ($I_{2\pi}(\phi)$) and Andreev bound states of STiS junction would be quite different
from those of the SNS junction.

However in earlier
work,\cite{Kitaev1,EPJB37-349(2004),PhysRevB.79.161408(2009)(L.Fu),PRL105-077001(2010),prb84-081304(2011),
PRL106-077003(2011),PRL107-177002(2011),JPCM24-325701(2012),PRL108-257001(2012),PRB86-140503(R)(2012),prb86-140504(R)(2012)}
only short STiS junction (junction length $L$ far less than the
superconductor coherent length $\xi_0$) is studied.
And it's only very recently we notice that the work by Beenakker
$et$ $al.$ \cite{Phys.Rev.Lett110-017003(2013)} discusses the finite
length clean junction. To the best knowledge of us, a study of the
finite length STiS junction affected by magnetic impurity is still
missing. That is the gap we want to fill here.

In this article both the $2\pi$-period and $4\pi$-period
current-phase relation is calculated. There is a sharp jump at
$\phi=\pi$ and $\phi=2\pi$ for $I_{2\pi}$ and $I_{4\pi}$
respectively in the clean junction. For $I_{2\pi}$, the sharp jump
at $\phi=\pi$ is robust against impurity strength and distribution.
However for $I_{4\pi}$, the impurity makes the jump at $\phi=2\pi$
smooth. The critical current and shape of current-phase
characteristics are greatly influenced by junction length.

The rest of the paper is organized as follows. In Sec.
\ref{sec2}, we describe the model and give the analytical results.
In Sec. \ref{sec3}, the numerical results and analysis are
given. In Sec. \ref{sec4}, we give a brief conclusion.
In Appendix A, we give the reason of the
similarity between STiS junction and conventional SNS junction. In
Appendix B and C, we derive the current operator and give the detail
of the calculation.

\section{Model and analytical results}
\label{sec2}

Two s-wave superconductors are in intimate contact with one edge of
2D TI. Because of the proximity effect, a 1D TS forms in the
interface. Then we have a STiS Josephson junction on one edge of the 2D
TI.\cite{txt3} The effective Hamiltonian of the edge state
is given as $H_0=v_F \sigma_3 p_x$, in which
$p_x=-i\hbar\partial_x$, $\sigma_{1,2,3}$ are Pauli matrices acting
in the spin space and $v_F$ is the velocity of the edge
states.\cite{review(Ti)(C.L.Kane)} Proximity effect contributes a
paring term, then the Hamiltonian of the 1D TS is given
as,\cite{PhysRevB.79.161408(2009)(L.Fu)}
\begin{eqnarray}
H=\int dx \psi^\dag  (H_0-\mu) \psi +\Delta \psi_\uparrow^\dag
\psi_\downarrow^\dag +\Delta^*\psi_\downarrow \psi_\uparrow
\end{eqnarray}
in which $\psi=(\psi_\uparrow,\psi_\downarrow)^T$, $\psi_\uparrow \
(\psi_\downarrow)$ annihilates a right (left)-moving electron.
$\Delta=\Delta_0 e^{i\phi^\prime}$ is the paring potential, $\Delta_0=|\Delta|$ and $\phi^\prime$ is the phase of the superconductor.
In Nambu representation $\Psi =
(\psi_\uparrow,\psi_\downarrow,\psi_\downarrow^\dag,-\psi_\uparrow^\dag)^T
$, with $i\hbar \partial_t \Psi=H_{BdG}\Psi$ we derive the
Bogoliubov-de Gennes (BdG) Hamiltonian\cite{PhysRevB.79.161408(2009)(L.Fu),book1(P.G.de Gennes)}
\begin{eqnarray}\label{eqBDG}
H_{BdG}=v_F  p_x \sigma_3\tau_3-\mu \tau_3 + \Delta_0[cos(\phi^\prime)\tau_1-sin(\phi^\prime)\tau_2],
\end{eqnarray}
where $\mu $ is the chemical potential and $\tau_{1,2,3}$ are Pauli matrices mixing the $\psi$ and $\psi^\dag$
blocks of $\Psi$. Particle hole symmetry is
expressed as $\{H_{BdG},\Xi\}=0$, in which $\Xi=\sigma_2\tau_2K$ and
K is the complex conjugation operator. As a result these states are not
independent. For an infinite TS the dispersion relation is $\epsilon=\pm
\sqrt{\hbar^2 v_F^2 (k\pm k_F)^2+|\Delta|^2}$, in which $\mu=\hbar v_F k_F$.
And we neglect the
self-consistency condition of $\Delta$.\cite{book1(P.G.de Gennes)}
For the junction considered here, $\Delta=\Delta_0
e^{i\phi_1}\theta(-x)+\Delta_0 e^{i\phi_2}\theta(x-L)$ where $L$ is
the length of the junction.

We include a region with magnetic impurity by adding a scattering term
$M\sigma_1\theta(x-L_1)\theta(L_2-x)$ in $H_{BdG}$. The magnetic
impurity can change the direction of particles, which can be
described by the scattering matrix for electrons and holes
\begin{eqnarray}\label{eqsca}
S_e=
\left(\begin{array} {lcr}
 r& t\\ t& -r^*t/t^*
\end{array}\right)
\quad and \quad
S_h=
\left(\begin{array} {lcr}
 -r^*& t^*\\ t^*& rt^*/t
\end{array}\right)
\end{eqnarray}
We denote the reflection coefficient $R=|r|^2$ and transition
coefficient $T=|t|^2$. For simplicity, we have assumed that $R$ is a
constant independent of energy and the length of the impurity region. Under this
assumption the effect of the length of the impurity region is
equivalent to replacing the junction length $L$ with an effective
length $L^\prime=L-(L_2-L_1)$, and in the following we abbreviate
$L^\prime$ to $L$. Comparing with SNS junction,\cite{txt2} there is an extra
$\pi$ phase shift for hole reflection, and that's the origin of
difference between STiS and SNS junction in Andreev bound states and $I_{2\pi}$ (see Appendix
A for an explanation).

Incident particles with energy $\epsilon$ will be reflected at the superconductor-normal
interface.\cite{JETP19-1228(1964)(Andreev)}
For SNS junction, it can occur both the Andreev and normal reflections at the interface.
But for STiS junction, only the quantum Andreev reflection occurs at the
interface.\cite{PRL109-186603(2012)(RRDu),Adroguer,sun1}
If $|\epsilon|<\Delta_0$, incident particles will be reflected completely, therefore Andreev bound states will form.\cite{JETP30-944(1970)(Kulik)}
Solve the BdG equation, then we obtain the energy level equation of Andreev bound states.
For clean junction
\begin{eqnarray}\label{eq1}
-2 arccos(\frac{\epsilon}{\Delta_0})+\frac{\epsilon}{\Delta_0}\frac{L}{\xi_0}=\pm \phi+2\pi n
\end{eqnarray}
where $\xi_0=\hbar v_F/(2\Delta_0)$ is the superconducting coherent
length, $\phi=\phi_2-\phi_1$ is the phase difference and
$n=0,\pm1,\pm2,...$ . The second term on the left side of
Eq.(\ref{eq1}) is equal to $(k_e-k_h)L$, where $k_e\,(k_h)$ is the
wave vector of the right-moving electron (left-moving hole) with energy
$\epsilon$. Then we can interpret Eq.(\ref{eq1}) in terms of
Bohr-Sommerfeld quantization of the periodic electron-hole orbits in
the TI region. \cite{book2(A.A.Abrikosov)} In the presence of
impurity Andreev bound state is given as
\begin{eqnarray}\label{eq2}
-2arccos(\frac{\epsilon}{\Delta_0})+\frac{\epsilon}{\Delta_0}\frac{L}{\xi_0}=\alpha
\end{eqnarray}
in which the phase difference is changed to $\alpha$,
\begin{eqnarray}
cos(\alpha)=Tcos(\phi)- R\, cos(\frac{\epsilon}{\Delta_0}\frac{L-2L_1}{\xi_0} )
\end{eqnarray}
which is different from that of the SNS
junction.\cite{PhysRevB.46.12573(1992)(P.F.Bagwell)}

The Josephson current $I(\phi)$ induced by the superconducting phase
contains two parts, the discrete current $I_d(\phi)$ and the continuous
current $I_c(\phi)$ carried by quasiparticles occupying Andreev bound states and
continuous energy spectrum respectively. To compute the current, we
suppose the system is nearly in thermodynamic equilibrium. Because
the current is constant, we can solve the wave function and then
obtain the average value of current operator in the TI region. The
current due to the scattering state (the eigenstate of junction Hamiltonian)
$\varphi=(u(x),u^\prime(x),v(x),v^\prime(x))^T$ with eigenvalue $\epsilon$ is
\begin{eqnarray}\label{eqco}
J=ev_F[(|u|^2+|v|^2-|u^\prime|^2-|v^\prime|^2)f(\epsilon)-|v|^2+|v^\prime|^2]
\end{eqnarray}
where $e$ is the electron charge and $f(\epsilon)$ is the Fermi distribution function.
The last two terms describe the current carried by the ``vacuum"
(spin-down band and spin-up band filled by electrons) on which we
can create quasiparticles occupying the ground state of $H_{BdG}$ to obtain the superconducting
ground state.\cite{PLDS3-1(1996)(S.Datta)} There is an alternative
statistical method by which current is the derivative of free
energy. In this article we use the wave function method to calculate
the continuous current and the quantum statistical method for the
discrete current. In appendix B and C we give the calculation detail and prove results according to both
methods are equivalent for the discrete current.

The discrete current can be written as $I_d(\phi)=\sum_n I_n(\phi)f(\epsilon_n)$,
where $I_n(\phi)$ is the current
carried by the quasiparticle occupying Andreev bound state with eigenvalue $\epsilon_n$.
According to the quantum statistical method, the effective current due to Andreev bound state with eigenvalue $\epsilon_n$ is $I_n(\phi)=\frac{e}{\hbar}\frac{d\epsilon_n}{d\phi}$ (derived in Appendix C).
For dirty junction,
\begin{eqnarray}
I_n(\phi)&=&\frac{1}{2}\frac{ev_F}{L +2\xi(\epsilon_n)}\frac{T sin(\phi)}{sin(\alpha)}\frac{1}{\gamma}\nonumber\\
\gamma&=&1+\frac{\hbar}{2e\Delta_0}\frac{ev_F}{L+2\xi(\epsilon_n)}\frac{R}{sin(\alpha)}\frac{L-2L_1}{\xi_0}\nonumber\\
&\times& sin(\frac{\epsilon_n(L-2L_1)}{\Delta_0 \xi_0})
\end{eqnarray}
For clean junction,
\begin{eqnarray}
I_j^\pm(\phi)=\pm \frac{1}{2} \frac{ev_F}{L +2\xi(\epsilon_j^\pm)}
\end{eqnarray}
where $\xi(\epsilon)=\xi_0\frac{\Delta_0}{\sqrt{\Delta^2-\epsilon^2}}$ is the energy dependent coherent length.

For a short junction ($L<<\xi_0$), it's enough to consider discrete current only,
because the continuous current is of the order of $L/\xi_0$. However for a long junction the continuous current can not be neglected.
To calculate $I_c(\phi)$, we first construct the scattering state for an incident particle having energy $\epsilon$,
and then apply the current formula given by Eq.(\ref{eqco}).
And we take the semiconductor picture (both the positive and negative solutions of BdG equation are used).
The detail of constructing scattering states and computing current is similar to Ref.\mycite{PhysRevB.46.12573(1992)(P.F.Bagwell)},
and some detail is given in Appendix B.
Results are given below.
For clean junction,
\begin{eqnarray}\label{eqc10}
I_c(\phi)&=&\frac{e}{h}T(\int_{-\infty}^{-\Delta_0}+\int_{\Delta_0}^{\infty} )d\epsilon f(\epsilon)|u_0^2-v_0^2| \nonumber\\
&\times&[\frac{1}{D(\epsilon,-\phi)}-\frac{1}{D(\epsilon,\phi)}]
\end{eqnarray}
For dirty junction,
\begin{eqnarray}\label{eqc11}
I_c(\phi)&=&\frac{e}{h}T\int d\epsilon f(\epsilon)|u_0^2-v_0^2|\frac{sin(\phi)}{sin(\alpha)}\nonumber\\
&\times&[\frac{1}{D(\epsilon,-\alpha)}-\frac{1}{D(\epsilon,\alpha)}]
\end{eqnarray}
in which
\begin{eqnarray}\label{eqc12}
D(\epsilon,\alpha)&=&u_0^4+v_0^4-2u_0^2v_0^2cos(\frac{\epsilon L}{\Delta_0 \xi_0}+\alpha)\nonumber\\
2u_0^2&=&1+\frac{\sqrt{\epsilon^2-\Delta_0^2}}{\epsilon}\nonumber\\
2v_0^2&=& 1-\frac{\sqrt{\epsilon^2-\Delta_0^2}}{\epsilon}, \quad
u_0v_0=\frac{\Delta_0} {2\epsilon}
\end{eqnarray}

\section{Numerical results and analysis}
\label{sec3}

For the short junction ($L<<\xi_0$), only a pair of Andreev bound states contributes to
current, and analytical result is available.\cite{PhysRevB.79.161408(2009)(L.Fu)} Andreev bound states are given as
$\epsilon=\pm \epsilon_0$, $\epsilon_0=\sqrt T \Delta_0 cos(\phi/2)$. The
corresponding current is $I=I_0 tanh(\frac{\epsilon_0}{2k_BT_B})$,
$I_0(\phi)=\frac{e}{2\hbar}\sqrt T \Delta_0sin(\phi/2)$ where $k_B$ and $T_B$ are Boltzmann constant and temperature respectively. However for
the finite length junction, we mainly give numerical analysis.

\begin{figure}[htbp]
\centering
\includegraphics[width=8cm,height=3.5cm]{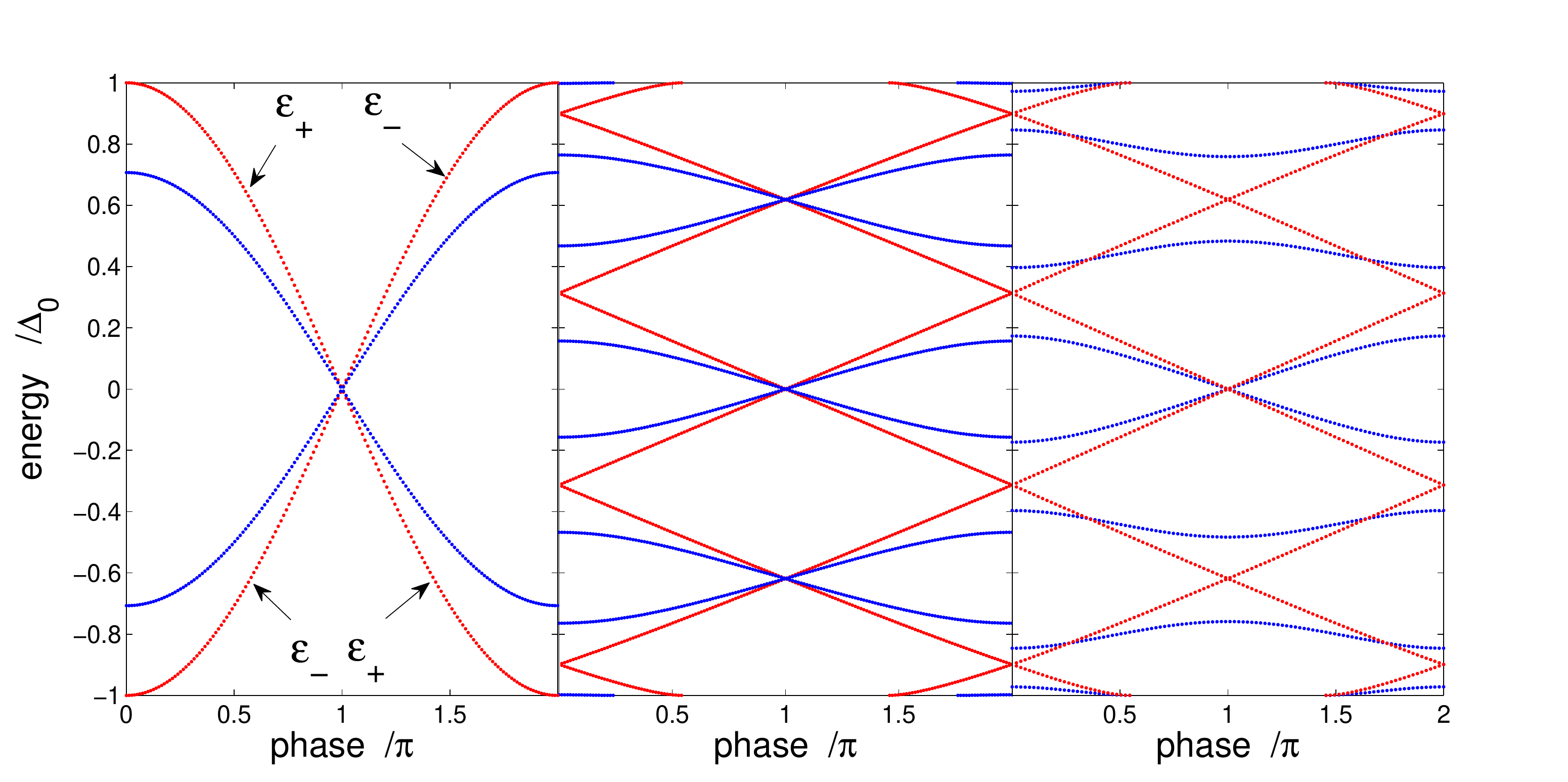}\\
\caption{(Color online) Andreev bound states with several junction length and transition coefficients.
Red line T=1, blue line T=0.5. Left: $L=0$. Middle: $L=8\xi_0$, $L1=L/2$. Right: $L=8\xi_0$, $L1=0.2L$.
}
\label{fig1}
\end{figure}

Fig.\ref{fig1} shows the effect of junction length and impurity
reflection on Andreev bound states. The length of the junction will increase the
number of bound states consistent with the usual 1D quantum wells.
The number can be given approximately as $2\,Int(L/(\xi_0\pi))+2$ or
$2\,Int(L/(\xi_0\pi))+4$, where $Int(x)$ means the integer part of
$x$. The symmetrical impurity ($L_1=L/2$) opens a gap at
$\phi=2n\pi$ ($n$ is the integer) as shown in the middle panel of
Fig.1. For the asymmetrical impurity ($L_1\neq L/2$), it can open
the gap at both $\phi=2n\pi$ and $\phi=(2n+1)\pi$ (see the right
panel of Fig.1). But the crossing point at $\phi=\pi,\epsilon=0$ remains
for arbitrary length and can not be broken by impurity scattering
which is different from the conventional SNS
junction.\cite{PhysRevB.46.12573(1992)(P.F.Bagwell)} That specific
crossing point is protected by the fermion parity
conversion.\cite{PhysRevB.79.161408(2009)(L.Fu)}
\begin{figure}[htbp]
\centering
\includegraphics[width=7.5cm]{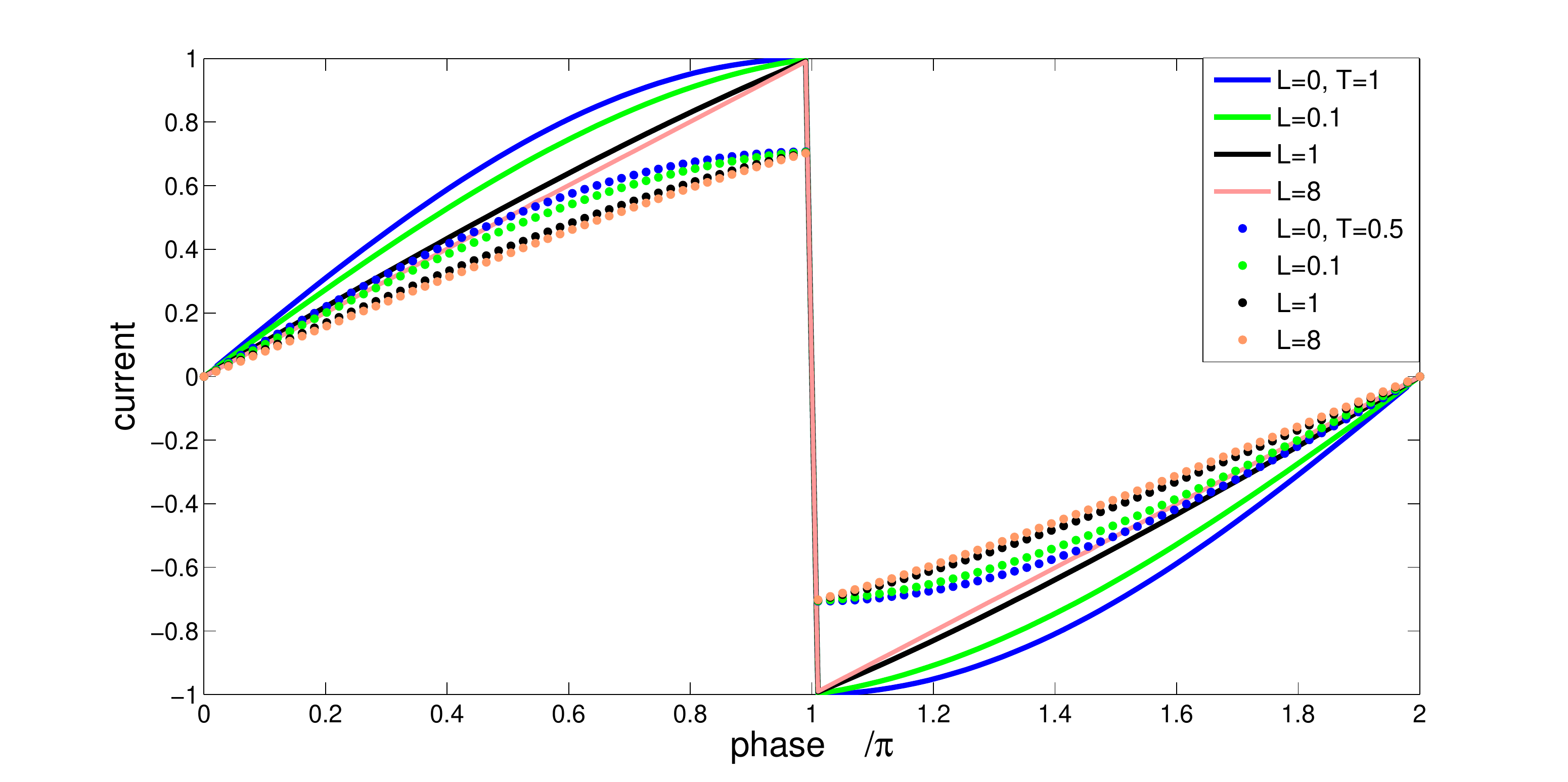}\\
\caption{(Color online) Current-phase relation for several junction
length with certain transition coefficient. T=1 (0.5) for solid (dotted) line. For dotted lines the impurity distributes symmetrically.
Current in unit of $\frac{ev_F}{2(L+2\xi_0)}$. Length in unit of $\xi_0$.
Temperature is zero. } \label{fig2}
\end{figure}

\begin{figure}[htbp]
\centering
\includegraphics[height=4cm]{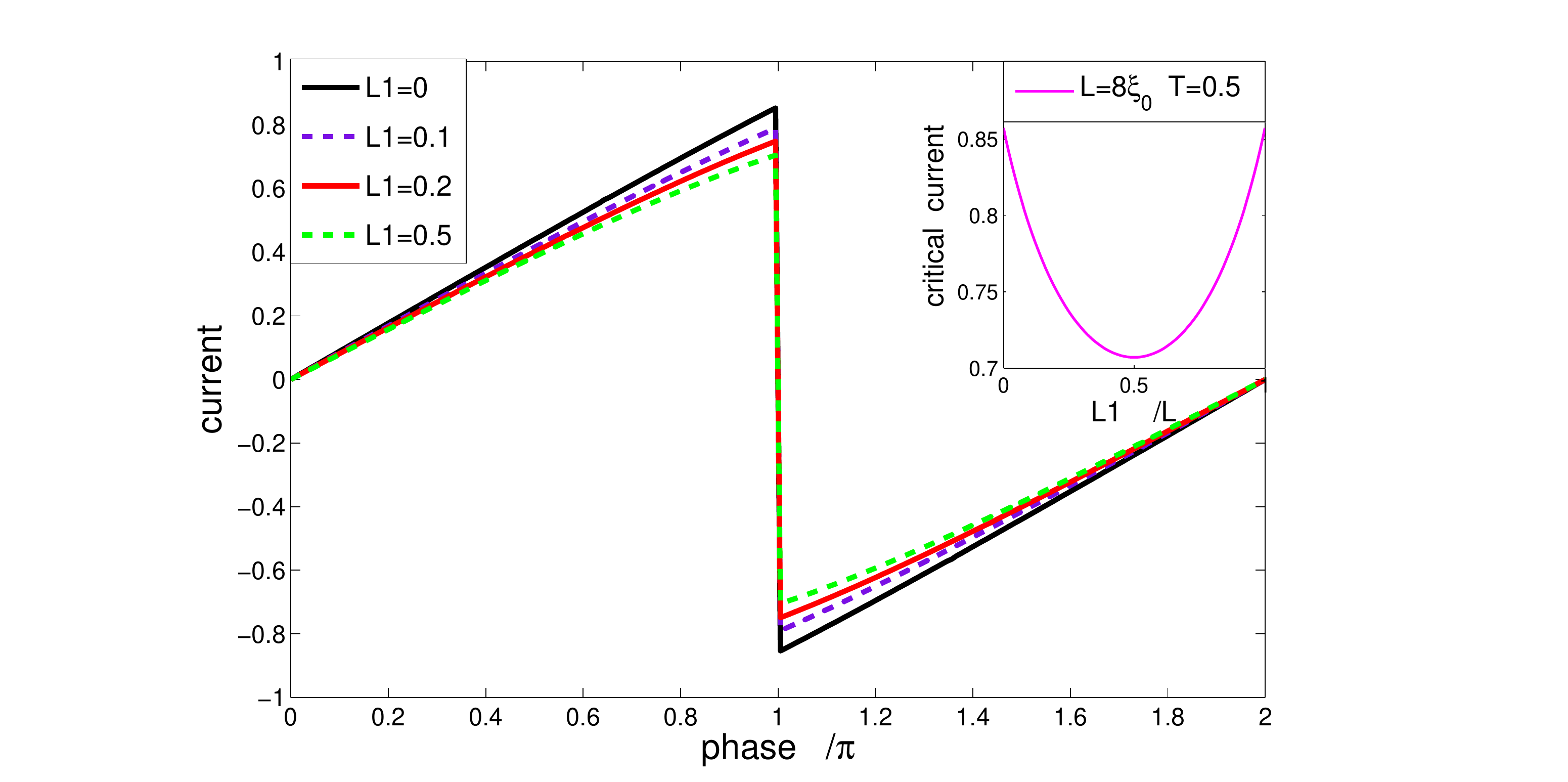}\\
\caption{(Color online) Current-phase relation for several L1.
Current in unit of $\frac{ev_F}{2(L+2\xi_0)}$. L1 in unit of L. $T=0.5$, $L=8\xi_0$.
Inset shows the dependence of critical current on L1.
Temperature is zero.
}
\label{fig3}
\end{figure}

The zero temperature current-phase characteristics for different
junction length and impurity strength and distribution are shown in
Fig.\ref{fig2} and Fig.\ref{fig3}. With the length increasing, the
curve changes from sinusoidal to sawtooth. Impurity reflection is mainly
to decrease the critical current. There is a robust sharp jump at $\phi=\pi$. Since the continous current
is zero while $\phi=\pi$, the jump is rooted in the
crossing point of Andreev bound state at $\phi=\pi, \epsilon=0$. It will not be destroyed
by impurity reflection because the impurity can not open a
gap at $\phi=\pi, \epsilon=0$, which is different from the case of
conventional SNS junction.

The critical (maximum) current $I_{c,2\pi}$ is reached when $\phi=\pi$, with
$I_{c,2\pi}=I_d(\pi)$ due to $I_c(\pi)=0$. For the dirty junction
\begin{eqnarray}
I_{c,2\pi}=\frac{ev_F}{2}\frac{\sqrt T}{\sqrt{(L+2\xi_0)^2-R(L-2L_1)^2}}
.\end{eqnarray}
For the clean junction $I_{c,2\pi}=\frac{1}{2}\frac{ev_F}{L+2\xi_0}$. For
symmetrical impurity distribution, $I_{c,2\pi}|_T=\sqrt T I_{c,2\pi}|_{T=1}$. In
this case, the impurity reflection monotonously decreases the
critical current. The asymmetrical impurity distribution will
enhance the current shown in the inset of Fig.\ref{fig3}. That's
different from the conventional SNS case where the critical current
will decrease when impurity leaves the center. For the long junction
with extremely asymmetrical impurity distribution ($L>>\xi_0, \
L>>L_1$), we have $I_{c,2\pi}\approx \frac{1}{2}\frac{ev_F}{L+2\xi_0}$ for
$T$ not too small, which nearly reaches the result of clean
junction.

In the previous discussion, we suppose that there is some mechanism
to make quasiparticles distribute nearly thermodynamically. Now
we consider the case that the necessary mechanism is absent for
the two eigenstates $\varphi_\pm(\phi)$ with energy $\epsilon_\pm(\phi)$
nearest to zero shown in Fig.\ref{fig1}. The two states are
connected by electron-hole symmetry, $\varphi_+=\Xi\varphi_-$ and
$\epsilon_-=-\epsilon_+$.

The original state remains
while phase difference is changed adiabatically. Starting from ground state while $\phi=0$, for $\phi<2\pi$ state
$\epsilon_-$ is occupied. The current due to a pair of Andreev bound states is
\begin{eqnarray}
I_e=\frac{e}{\hbar}\frac{\partial\epsilon_-}{\partial\phi}f(\epsilon_-)
-\frac{e}{\hbar}\frac{\partial\epsilon_-}{\partial\phi}(1-f(\epsilon_-))
\end{eqnarray}
and the distribution is $f(\epsilon_-)=1$ independent of energy, then
we have $I_e=\frac{e}{\hbar}\frac{\partial\epsilon_-}{\partial\phi}$
for $0<\phi<2\pi$. While $\phi=2\pi$, the system is in
excited state. And it can not decay to ground state because of
fermion parity conversion.\cite{PhysRevB.79.161408(2009)(L.Fu),RPP75.076501(2012)(J.Alicea)} For $2\pi<\phi<4\pi$, the state
$\epsilon_+$ is occupied. While $\phi=4\pi$,
the system reaches the original state we start
with.\cite{PhysRevB.79.161408(2009)(L.Fu),RPP75.076501(2012)(J.Alicea)}
Therefore $I_e$ is $4\pi$ periodic. The net current will be $4\pi$
periodic since $I_e$ contributes significantly to current.

\begin{figure}[htbp]
\centering
\includegraphics[height=4.2cm]{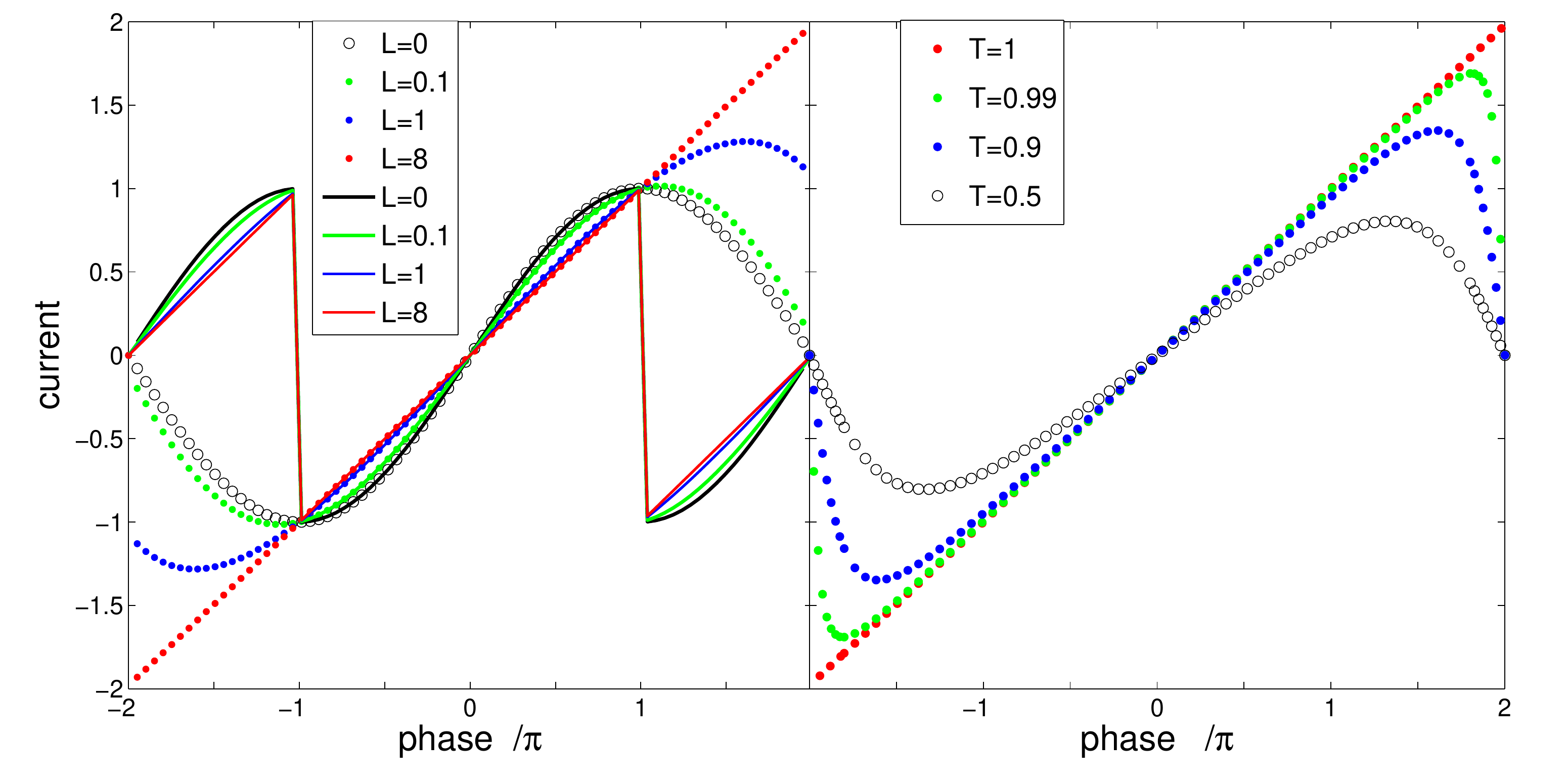}\\
\caption{(Color online) Current-phase relation to show the $4\pi$
period. Dotted (solid) line for $I_{4\pi}$ ($I_{2\pi}$). Left: T=1.
Right: L=$8\xi_0$, $L_1=0.5L$. Current in unit of
$\frac{ev_F}{2(L+2\xi_0)}$. Length in unit of $\xi_0$. Temperature is zero. } \label{fig4}
\end{figure}

The current-phase curve is shown in Fig.\ref{fig4}. There
is a sharp jump at $\phi=2\pi$ for $I_{4\pi}$ in finite length clean junction.
For $I_{2\pi}$, the jump at $\phi=\pi$ is robust against impurity reflection.
However impurity reflection will make the jump located at
$\phi=2\pi$ smoother for $I_{4\pi}$. The reason is that for clean
junction the energy crossing of Andreev bound state at $\phi=2\pi$ has a none-zero
slope. While for the dirty junction the slope is zero (see Fig.\ref{fig1}). Here we
denote the maximum of $I_{4\pi}$ $(I_{2\pi})$ as $I_{c,4\pi}=gI_{c,2\pi}$.
$g$ increases with length increasing.
We have $g=1$ for the junction with length $L=0$. For the long clean junction ($L>>\xi_0$),
$g=2$.\cite{Phys.Rev.Lett110-017003(2013)} That's apparent if we
notice that the energy level located deeply in the paring potential
well is nearly linear for the long clean junction. Impurity reflection
will make the factor decrease. For long junction case, vary the reflection coefficient
from 0 to 1, $g$ changes from 2 to 1. For a short junction ($L=0$),
$g$ is independent of reflection and we have $g=1$.

\section{Conclusion}
\label{sec4}

In summary, the current-phase relation of a finite length STiS
junction with magnetic impurity is investigated. We consider both
the $2\pi$- and $4\pi$-period case. With the length increasing, the
current-phase curve evolves form a sinusoidal shape into
sawtooth shape. There is a sharp jump at $\phi=\pi$ and $\phi=2\pi$
for $I_{2\pi}$ and $I_{4\pi}$ respectively in the clean junction.
For $I_{2\pi}$, the sharp jump at $\phi=\pi$ is robust against
impurity strength and distribution. However for $I_{4\pi}$, the
impurity makes the jump at $\phi=2\pi$ smooth.
The critical current is greatly influenced by junction length and impurity.
\section*{ACKNOWLEDGMENTS}

This work was financially supported by NBRP of China (2012CB921303
and 2009CB929100) and NSF-China under Grants Nos. 11074174 and
11274364. One of the authors (S.F.Zhang) wishes to thank H.W.Liu and
S.N.Zhang for helpful discussions.

\section*{Appendix A}
This appendix explains the origin of the similarity and difference between the STiS junction and conventional SNS junction.

For the SNS junction, Nambu basis can be selected in two equivalent
form, $\psi_+=(\psi_\uparrow,\psi_\downarrow^\dag)^T$ or
$\psi_-=(\psi_\downarrow,\psi_\uparrow^\dag)^T$, due to spin
degeneracy. With $i\hbar\partial_t\psi_\pm=H_{BdG}\psi_\pm$, we can
derive $H_{BdG}=(p_x^2/2m
-\mu)\tau_3+\Delta_0[cos(\phi^\prime)\tau_1-sin(\phi^\prime)\tau_2]$,
$p_x=-i\hbar\partial_x$ and $m$ is the effective mass of electron. Take the Andreev
approximation\cite{txt5,JETP19-1228(1964)(Andreev)} and denote the
eigenvector as $\varphi=\chi e^{i(\sigma k_F+\delta k) \, x}$,
$\sigma=\pm$ for incident particles with wave vector near $\pm
k_F$, $\chi$ is a vector independent of $x$. Then we arrive at the
Andreev equation\cite{JETP19-1228(1964)(Andreev)}
\begin{eqnarray}
\left(\begin{array} {cc}
\sigma v_F p_x-\hbar v_F k_F & \Delta \\ \Delta^* & -\sigma v_F p_x+\hbar v_F k_F \end{array}
\right) \varphi=\epsilon\,\varphi
\end{eqnarray}
If we reset Nambu basis as
$\Psi^\prime=(\psi_\uparrow,\psi_\downarrow,\psi_\downarrow^\dag,\psi_\uparrow^\dag)^T$
and take $\sigma=1(-1)$ for $\psi_+$($\psi_-$), we will find the
corresponding BdG Hamiltonian is identical to the BdG Hamiltonian of
STiS junction, but the Nambu basises are connected with a unitary
transformation $ P= \left(\begin{array} {lr} \sigma_0&\\ & \sigma_3
\end{array} \right) $, in which $\sigma_0$ is a $2\times2$ unit
matrix. $P$ matrix leads that for dirty STiS junction there will be
an extra $\pi$ phase shift for hole reflection as is shown in
Eq.(\ref{eqsca}). The other choice to take $\sigma=1(-1)$ for
$\psi_-$($\psi_+$) corresponds to the same junction formed on the
other side of the 2D TI.

\section*{Appendix B}
\label{sec5}

This appendix is to derive the current formula
Eq.(\ref{eqco}) and give some detail of calculating the current.

\begin{figure}[htbp]
\centering
\includegraphics[height=4cm]{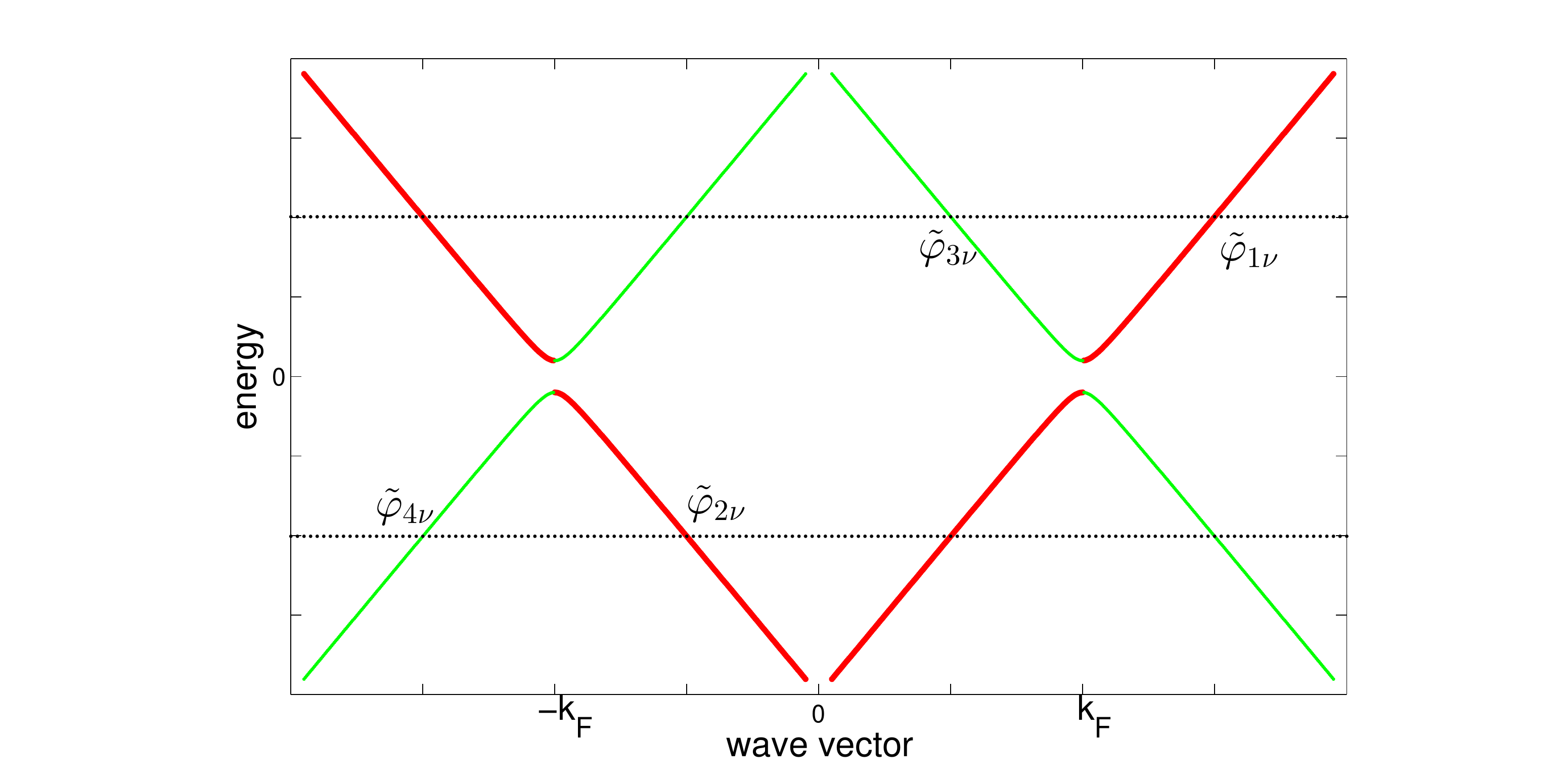}\\
\caption{(Color online) Continuous energy spectrum for infinite 1D TS. Red and green lines
correspond to electronlike and holelike eigenstates of 1D TS respectively. $\tilde\varphi_{i\nu}$ is the the incident state from which one can construct the scattering state $\varphi_{i\nu}$.
} \label{figa1}
\end{figure}

The system is given as
\begin{eqnarray}
H&=&\int dx\psi^\dag (H_0-\mu) \psi +\Delta \psi_\uparrow^\dag
\psi_\downarrow^\dag +\Delta^*\psi_\downarrow \psi_\uparrow
\end{eqnarray}
With $i\hbar \partial_t \Psi=H_{BdG}\Psi$, the BdG Hamiltonian is yielded
\begin{eqnarray}\label{HBdG}
H_{BdG}=\left(\begin{array}{cc} H_0 -\mu & \Delta\\ \Delta^* &-\hat
T (H_0-\mu) \hat T^{-1} \end{array}\right)
\end{eqnarray}
with time reversal operator $\hat T=-i\sigma_2 K$. In fact
Eq.(\ref{HBdG}) is appropriate for arbitrary $H_0$ but
with the corresponding time reversal operator for different systems.
The BdG equation can be written as
\begin{eqnarray}
 H_{BdG}\;\varphi_{i,\nu}(x)=\epsilon_{i,\nu}\, \varphi_{i,\nu}(x)
\end{eqnarray}
where $\varphi_{i,\nu}=(u_{i,\nu}(x),u_{i,\nu}^\prime(x),v_{i,\nu}(x),v_{i,\nu}^\prime(x))^T$
is the eigenvector and $\epsilon_{i,\nu}$ is the eigenvalue.
Because of electron-hole symmetry $\{H_{BdG},\Xi\}=0$,
$\Xi\varphi_{i,\nu}(x)$ is also an eigenvector with eigenvalue
$-\epsilon_{i,\nu}$. $\nu$ and $i$ denote energy and the extra
degeneracy respectively. For continuous spectrum $i=1,2,3,4$,
$\varphi_{3,\nu}=\Xi \varphi_{2,\nu}, \varphi_{4,\nu}=-\Xi \varphi_{1\nu}$.
$\varphi_{i,\nu}$ is the scattering state constructed from the incident state (eigenstate of 1D TS) $\tilde\varphi_{i,\nu}$ shown in
Fig.\ref{figa1}. However for Andreev bound states we only have $i=\pm$,
$\varphi_{-,\nu}=\Xi\varphi_{+,\nu}$. For simplicity we denote
$\varphi_{-,\nu}=\varphi_{1,\nu}$, $\varphi_{+,\nu}=\varphi_{4,\nu}$ and
$\varphi_{2,\nu}=\varphi_{3,\nu}=0$.

To diagonalize the Hamiltonian we first rewrite it as
$H=\frac{1}{2}\int dx\Psi^\dag H_{BdG}\Psi+constant$. The Bogoliubov
transformation is given as $\Psi=\sum_\nu S_\nu \gamma_\nu$, in
which $S_\nu=(\varphi_{1,\nu},\varphi_{2,\nu},\varphi_{3,\nu},\varphi_{4,\nu})$,
$\gamma_\nu=(\gamma_{1,\nu},\gamma_{2,\nu},\gamma_{3,\nu},\gamma_{4,\nu})^T$,
$\gamma_{4,\nu}=-\gamma_{1,\nu}^\dag$,
$\gamma_{3,\nu}=\gamma_{2,\nu}^\dag$. $\gamma_{i\nu}$ annihilates a
quasiparticle in eigenstate $\varphi_{i\nu}$. Then we have
$H=\frac{1}{2}\sum_{i,\nu}\epsilon_{i,\nu}\gamma_{i,\nu}^\dag\gamma_{i,\nu}+constant$.

The current density operator can be derived with the current density
conversion equation, $\partial_t \hat\rho(x)+\partial_x\hat J(x)=0$,
in which electron density operator
$\hat\rho(x)=e(\psi_\uparrow^\dag(x)\psi_\uparrow(x)+\psi_\downarrow^\dag(x)\psi_\downarrow(x)
)$. In the TI region, it can be derived as $\hat
J(x)=ev_F(\psi_\uparrow^\dag(x)\psi_\uparrow(x)-\psi_\downarrow^\dag(x)\psi_\downarrow(x))$.
In the TS region, the paring potential will contribute an additional term $-\partial_x
\hat
J_s=2e(\Delta\psi_\uparrow^\dag\psi_\downarrow^\dag-H.C.)/i\hbar$,
which describes exchanging Cooper pairs between quasiparticles and
condensate. However this term vanishes for energy larger than paring
potential, thus it makes no contribution to the continuous current.
But it will make the discrete current transforms into supercurrent
carried by the condensate gradually in the superconducting region.
\cite{PhysRevB.25.4515(1982)(BTK)}

Take the ensemble average $J(x)=\ <\!\!\hat J(x)\!\!>$, with
Bogoliubov transformation and
$<\!\!\gamma_{i,\nu}^\dag\!\gamma_{i,\nu}\!\!> \
=f(\epsilon_{i,\nu})$. In the TI region we find $J=\sum_\nu
J_{1,\nu}+J_{2,\nu}$,
\begin{eqnarray}\label{eqic}
J_{i,\nu}=&ev_F& [|u_{i,\nu}|^2f(\epsilon_{i,\nu})-|v_{i,\nu}|^2(1-f(\epsilon_{i,\nu})) ]\nonumber\\
&-&[|u_{i,\nu}^\prime|^2f(\epsilon_{i,\nu})-|v_{i,\nu}^\prime|^2(1-f(\epsilon_{i,\nu}))]
\end{eqnarray}
which is just Eq.(\ref{eqco}) we want to derive.
The extra current owing to paring potential is
$<\!\!-\partial_x\hat J_s\!\!> \ =\frac{4ev_F}{\hbar} Im \{\Delta\sum_{i=1,2;\nu}[u_{i,\nu}^* v_{i,\nu}
f(\epsilon_{i,\nu})-u_{i,\nu}^{\prime*} v_{i,\nu}^\prime f(\epsilon_{i,\nu})]\}$.

Now we prove that the contributions from electronlike and holelike injected states are equal.
$J_{e,\nu}$ ($J_{h,\nu}$) is the current due to electronlike (holelike) state $\varphi_{e,\nu}$ ($\varphi_{h,\nu}$) with eigenvalue $\epsilon_{e,\nu}$ ($\epsilon_{h,\nu}$) where $e=\{1,2\}$, $h=\{3,4\}$.
$J_{e/h,\nu}$ is given by Eq.(\ref{eqic}).
Since $\varphi_{h,\nu}=\Xi \varphi_{e,\nu}$,
$\epsilon_{h,\nu}=-\epsilon_{e,\nu}$ and
$1-f(\epsilon)=f(-\epsilon)$, we can obtain $J_{e,\nu}=J_{h,\nu}$.

For continuous spectrum, the eigenstate with a certain energy is 4-fold
degenerate. The continuous current can be written as, $I_c=\int
d\epsilon_\nu N(\epsilon_\nu) J(\epsilon_{\nu})/2$,
$J(\epsilon_\nu)=\sum_i J_i(\epsilon_{i\nu})=ev_F[\sum_{i}
(|u_{i,\nu}|^2+|v_{i,\nu}|^2-|u_{i,\nu}^\prime|^2-|v_{i,\nu}^\prime|^2)
f(\epsilon_{i,\nu})-\sum_i
|v_{i,\nu}|^2+\sum_i|v_{i,\nu}^\prime|^2]$, $N(\epsilon_\nu)$ is the
density of states of TS. Solve the eigenvectors and we find the last
two terms cancel with each other. Then we have $I_c=-\frac{1}{2}\int d\epsilon_{\nu}
N(\epsilon_{\nu})
[J_1^\prime(\epsilon_{\nu})+J_2^\prime(\epsilon_{\nu})]tanh(\frac{\epsilon_{\nu}}{2k_BT_B})
$, where $J_i^\prime(\epsilon_{\nu})=ev_F
(|u_{i,\nu}|^2+|v_{i,\nu}|^2-|u_{i,\nu}^\prime|^2-|v_{i,\nu}^\prime|^2)
$, which is the equation we use to derive Eqs. (\ref{eqc10})-(\ref{eqc12}).

\section*{Appendix C}

This appendix is to prove the discrete current obtained by wave function method
and quantum statistical method is identical if the states are
occupied thermodynamically. \cite{LTP23.181(1997)(G.Wendin)}

In this appendix we take the Nambu basis given as $\Psi^\prime=(\psi_\uparrow,\psi_\downarrow^\dag,\psi_\downarrow,-\psi_\uparrow^\dag)^T$ for simplicity.
The corresponding BdG Hamiltonian is
\begin{eqnarray}
H_{BdG}^\prime=
\begin{pmatrix}
(v_F p_x-\mu)\tau_3+\hat\Delta & M(x)\tau_0 \\
 M(x)\tau_0  & (-v_F p_x-\mu)\tau_3+\hat\Delta
\end{pmatrix}
\end{eqnarray}
where $M(x)=M\theta(x-L_1)\theta(L_2-x)$, $\tau_0$ is a $2\times2$ unit matrix and
 $ \hat\Delta= \left(\begin{array} {lr} & \Delta_0 e^{i\phi sgn(x)/2}\\\Delta_0 e^{-i\phi sgn(x)/2} &
\end{array} \right) $, $sgn(x)=x/|x|$.

A pair of Andreev bound states connected by electron-hole transformation is given as
$\varphi^\prime_\pm$ with energy $\epsilon_\pm$, $\varphi_+^\prime=\Xi\varphi_-^\prime$,
$\varphi_\pm^\prime=(u_\pm(x),v_\pm(x),u^\prime_\pm(x),v^\prime_\pm(x))^T$ and
$\epsilon_-=-\epsilon_+$. The corresponding current is
$J=ev_F(|u_-|^2+|v_-|^2-|u^\prime_-|^2-|v^\prime_-|^2)
(f(\epsilon_-)-f(\epsilon_+))/2+ev_F(|u_-|^2-|v_-|^2-|u^\prime_-|^2+|v^\prime_-|^2)/2$.
With the solved eigenvectors, we find the second term on the right side vanishes. Then the
current is derived as
\begin{eqnarray}\label{eqabs}
J=I(\epsilon_-)f(\epsilon_-)+I(\epsilon_+)f(\epsilon_+),
\end{eqnarray}
in which
$I(\epsilon_\pm)=ev_F(|u_\pm|^2+|v_\pm|^2-|u^\prime_\pm|^2-|v^\prime_\pm|^2)/2$
and $I(\epsilon_+)=-I(\epsilon_-)$. $I(\epsilon_\pm)$ can be seen as
the effective current carried by eigenstate $\varphi_\pm^\prime$.

Rewrite the current as $J=I(\epsilon_-)(f(\epsilon_-)-f(\epsilon_+))$,
and then act the operator $p_x=-i\hbar \partial_x$ on both sides.
With a straightforward calculation,
we have $2p_x J/e=\varphi_-^{\prime \dag} \begin{pmatrix} [\hat\Delta,\tau_3]&\\&[\hat\Delta,\tau_3]\end{pmatrix}\varphi_-^\prime\cdot(f(\epsilon_-)-f(\epsilon_+))$.
With $[\hat\Delta,\tau_3]=4i sgn(x) \frac{\partial \hat\Delta}{\partial\phi}$,
we derive
$\frac{-\hbar}{2e} sgn(x) \partial_x J=\varphi_-^{\prime\dag}\frac{\partial
H_{BdG}}{\partial \phi} \varphi_-^\prime (f(\epsilon_-)-f(\epsilon_+))$. Integrate among the the
whole region. Since the current in the TI region is
constant and it decays to zero gradually in the superconductor, the left side gives $\hbar J(x=0)/e$. With
the help of Feynman-Hellmann theorem the right side gives $\frac{\partial\epsilon_-}{\partial\phi}(f(\epsilon_-)-f(\epsilon_+))$.
Then we obtain
\begin{eqnarray}
J=\frac{e}{\hbar}\frac{\partial\epsilon_-}{\partial\phi}f(\epsilon_-)
+\frac{e}{\hbar}\frac{\partial\epsilon_+}{\partial\phi}f(\epsilon_+).
\end{eqnarray}
Comparing with Eq.(\ref{eqabs}) we have
$I(\epsilon_\pm)=\frac{e}{\hbar}\frac{\partial\epsilon_\pm}{\partial
\phi}$. So far we have proved the two methods are equivalent for the
discrete current.

It's of importance to point out that the quantum statistical method
has taken both particle energy levels and hole energy levels into
consideration. For a clean STiS junction with length $L=0$, one can take the Nambu basis to be
$\psi=(\psi_\uparrow, \psi_\downarrow^\dag )^T$ and the
corresponding BdG Hamiltonian is $2\times 2$. In this case there is
only one energy level with energy $\epsilon_-(\phi)$ contributes to current. However as we have discussed a
wrong result,
$J=\frac{e}{\hbar}\frac{\partial\epsilon_-}{\partial\phi}f(\epsilon_-)$,
will be derived if we use the quantum statistical method .


\begin{references}
\bibitem{txt1}
For Andreev bound states there is a difference of degeneracy. For $I_{2\pi}(\phi)$ there is a difference of factor 2.
\bibitem{txt3}
The effect of the other side is neglected because of  large space difference.
\bibitem{txt5}
Andreev approximation means $\mu>>|\Delta|$, the dispersion curve is approximated to be linear near $\pm k_F$.
\end{references}
\end{document}